\documentstyle[multicol,prb,aps,graphicx,amssymb]{revtex}
\setkeys{Gin}{width=0.9\linewidth}

\begin{document}
\draft

\title{Reentrance of the induced diamagnetism in gold-niobium 
proximity cylinders}

\author{F. Bernd M{\"u}ller-Allinger and Ana Celia Mota}
\address{Laboratorium f{\"u}r Festk{\"o}rperphysik, Eidgen{\"o}ssische 
         Technische Hochschule Z{\"u}rich, \\ 
8093 Z{\"u}rich, Switzerland} 
\date{\today}
\maketitle

\begin{abstract}
{\footnotesize We discuss the magnetic response of gold-niobium 
proximity cylinders in the temperature range $100\,\mu{\mathrm 
K}<T<9\,{\mathrm K}$.  The temperature dependence of the 
susceptibility in the diamagnetic regime is described well by the 
quasiclassical Eilenberger theory including an elastic mean-free path 
$\ell$.  In the mesoscopic regime, the temperature dependence of the 
reentrant paramagnetic susceptibility $\chi_{\mathrm 
para}(T)\propto\exp{[-L/3\xi^d_{N}(T)]}$ is governed by the dirty 
limit coherence length $\xi^d_{N}(T)=\sqrt{1/3\xi_{N}\ell_{N}}$, with 
$\xi_{N}$ the clean limit coherence length obtained from 
breakdown field measurements and $\ell_{N}$ the measured mean-free path 
in the gold layer.  At $T=100\,\mu{\mathrm K}$, $\chi_{\mathrm para}$ 
compensates about 1/5 of the induced diamagnetic susceptibility in gold.  }
\end{abstract}

\begin{multicols}{2}
\narrowtext The field of mesoscopic physics, including superconducting 
structures in proximity with normal metals, has attracted a wide 
interest, recently\cite{review}.  In particular, experimental work on 
transport properties of mesoscopic proximity systems has been 
enabled by recent advances in nanostructure fabrication 
technology.  In addition, extensive theoretical and 
experimental work on magnetic properties has led to the understanding 
of the high-temperature diamagnetic response of rather clean normal 
metal-superconductor (NS) proximity structures\cite{bmueller1} in the 
context of the quasiclassical Eilenberger theory\cite{belzig} as well 
as the magnetic breakdown transitions\cite{fauchere}.

On the other hand, the paramagnetic reentrance phenomenon, discovered 
some years ago in silver-niobium cylinders\cite{visaniprl} is still a 
matter of discussion.  Recently, three theoretical 
works\cite{bruder,fauchere99,maki} have addressed the origin of 
paramagnetic currents in NS systems from different points of view.  
Bruder and Imry\cite{bruder} consider non Andreev reflecting 
trajectories at the outer surface of NS proximity cylinders (glancing 
states), which give a paramagnetic correction to the Meissner 
susceptibility.  This result has been subject to debate because of its 
small magnitude\cite{comment}.  The approach by Fauch\`ere \textit{et 
al}.\cite{fauchere99} assumes a net repulsive interaction in the noble 
metals.  In their case, the $\pi$--shift of the order parameter then 
leads to spontaneous paramagnetic currents at the NS interface.  In 
the most recent work by Maki and Haas\cite{maki} the noble metals 
copper, silver, and gold are assumed to show p-wave superconducting 
ordering below transition temperatures close to where the paramagnetic 
reentrance effect occurs.  In their model, a quantized counter-current 
is generated in the periphery of the NS cylinders, compensating the 
Meissner effect.  Unfortunately, neither of the present 
theories\cite{bruder,fauchere99,maki} correctly explain the absolute 
value or the temperature dependence of the paramagnetic reentrant 
effect.  A more detailed discussion can be found in Ref.\ 
\onlinecite{bmueller2}.

In view of the most recent two theories\cite{fauchere99,maki}, which 
depend on specific (assumed) properties of the normal metals, it is 
important to check the reentrant effect in NS junctions of different 
materials.  Here, we report on the reentrance of the magnetic 
susceptibility of a gold-niobium sample, covering a large mesoscopic 
regime down to $\mu$K-temperatures.  This new material combination 
gold-niobium, follows our previous investigations\cite{bmueller2} of 
the paramagnetic reentrant effect in AgNb and CuNb specimens 
superimposing on fully induced Meissner screening.

The sample reported here is an ensemble of cylindrical wires with a 
superconducting core of soft niobium ($RRR\approx 300$) concentrically 
embedded in a normal-metal matrix of gold\cite{goodfellow}.  The total 
diameter was mechanically reduced by several steps of swagging and 
co--drawing \cite{flukiger} to the final value $23\,\mu{\mathrm m}$, 
with average normal layer thickness $3.2\,\mu{\mathrm m}$.  We chose 
not to anneal this sample, since gold dissolves in niobium as well as 
Nb in Au up to 5-10 percent for temperatures above 500\,$^\circ\mathrm 
C$\cite{roeschel}.  For this sample the value of the mean-free path 
$\ell_{N}\sim0.3d_{N}$ was obtained from resistivity measurements.  
The detailed sample preparation was done as for the samples 
reported in Ref.\ \onlinecite{bmueller1}.

A bundle of about 200 wires was placed directly inside the mixing 
chamber of a dilution refrigerator in contact with the liquid 
${\mathrm{^3He}}$--${\mathrm{^4He}}$ solution.  Using an rf--SQUID 
sensor, we measured inductively the temperature and field dependent 
ac-magnetic susceptibility as well as dc magnetization curves for 
$7\,{\mathrm mK}<T<7\,{\mathrm K}$.  For thermometry, the Curie--type 
magnetic susceptibility of the paramagnetic salt CMN (cerium magnesium 
nitrate) was measured, calibrated with Ge resistors.

Extensions of the measurements to $\mu$K temperatures were performed 
at the ultralow temperature (ULT) facility at the University of 
Bayreuth.  There, an experimental setup was installed for inductive 
measurements using an rf--SQUID sensor.  Magnetic fields were applied 
along the axis of the wires.

For the ULT experiments, we took parts of the wire bundle measured in 
our dilution refrigerator, and glued them with GE 7031 varnish to high 
purity gold foils tightly attached to a silver finger, in good 
electrical contact with the Cu demagnetization stage \cite{gloos88}.  
Thus, about 200 wires were mounted.  Temperatures were measured with a 
pulsed NMR Pt thermometer \cite{gloos88}.

\begin{figure}
\includegraphics[width=0.96\linewidth]{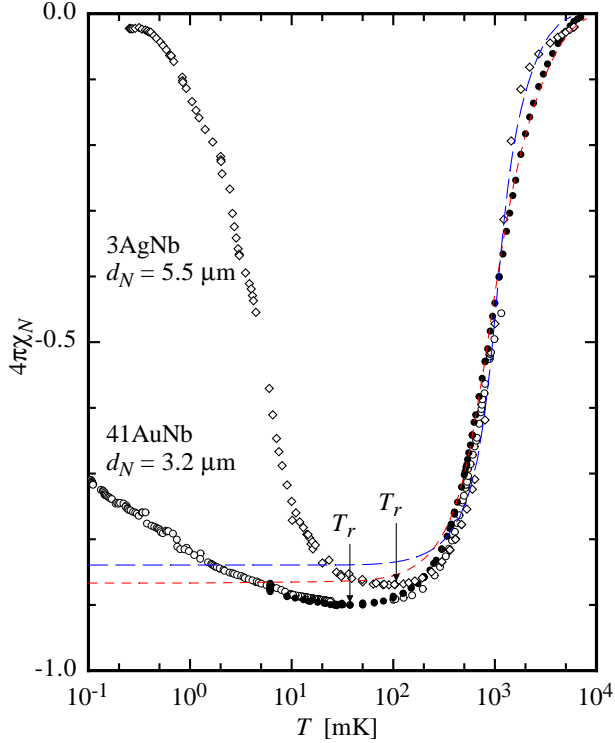}
		\caption{Magnetic susceptibility $\chi_{ac}$ ($\bullet$), 
		$\chi_{dc}$ ($\circ$), and numerical fit\protect\cite{belzig} 
		with $\ell/d_{N}=0.14$ (short-dashed line) of gold-niobium 
		sample~41AuNb as a function of temperature 
		($H_{dc}=0.2\,{\mathrm Oe}$).  For comparison $\chi_{ac}$, 
		$\chi_{dc}$ ($\diamond$), and numerical 
		fit\protect\cite{belzig} with $\ell/d_{N}=0.55$ (long-dashed 
		line) of silver-niobium sample~3AgNb.  }
    \protect\label{X(logT)41/3}
\end{figure}

In the following we report on the temperature dependent magnetic 
susceptibility of the gold--niobium sample~41AuNb [Fig.\ 
\ref{X(logT)41/3}].  We show $\chi_{ac}(T)$ between $7\,{\mathrm mK}$ 
and $9\,{\mathrm K}$, measured in our dilution refrigerator with field 
amp-
\noindent litude $H_{ac}=33\,{\mathrm mOe}$ ($H_{dc}\approx 0$) and 
frequency $\nu=80\,{\mathrm Hz}$, as well as $\chi_{dc}(T)$ at 
constant $H_{dc}= 0.2\,\mathrm{Oe}$ at ULT and LT.

Below the critical temperature of Nb ($T_{c}=9.2\,{\mathrm K}$), the 
magnetic susceptibility of the N layer exhibits diamagnetism induced 
through Andreev reflection at the NS interface.  At lower 
temperatures, it develops almost total Meissner screening in the Au 
layer, as in comparable AgNb samples \cite{bmueller1}.

The numerically obtained curve\cite{belzig} with only one parameter 
$\ell$ fits the experimental data very well over the whole temperature 
range above the Andreev temperature $T_{A}=\hbar v_{F}/2\pi 
k_{B}d_{N}$.  For sample~41AuNb it is $T_{A}=530\,\mathrm mK$.  The 
value of $\ell/d_{N}$ is $0.14$, about a factor of two smaller than 
the value $\ell_{N}/d_{N}=0.3$, obtained from resistivity 
measurements.

The fit is surprisingly good, if one takes into account the difference 
in geometry between experiment and theory, as well as the neglect of 
the boundary roughness in the theory and other imperfections 
inevitably present in the sample.  The quasiclassical theory of the 
proximity effect with a finite mean-free path parameter $\ell$ due to 
a low concentration of elastic scatterers\cite{belzig}, is able to 
describe the linear susceptibility data of our relatively clean AuNb 
specimen as well as it was previously shown for AgNb and CuNb 
systems\cite{bmueller1}.
\begin{figure}
\includegraphics[width=0.96\linewidth]{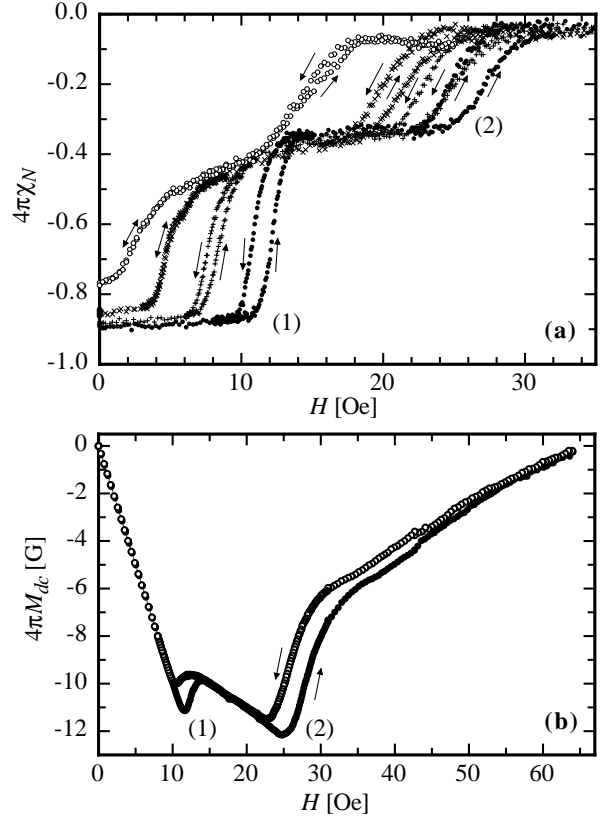}
		\caption{(a) Isothermal magnetic susceptibility $\chi_{ac}(H)$ 
		at temperatures $T=7\,\mathrm{mK}$ ($\bullet$), 
		$100\,\mathrm{mK}$ ($+$), $200\,\mathrm{mK}$ ($\times$), and 
		$400\,\mathrm{mK}$ ($\circ$) and (b) isothermal 
		dc-magnetization curve at $T=7\,\mathrm{mK}$ for 
		sample~41AuNb.  The low-(high-)field transitions correspond to 
		effective thickness $d^{(1)}_{N}=4.0\,\mu\mathrm{m}$ 
		($d^{(2)}_{N}=1.8\,\mu\mathrm{m}$) of the gold layer.  The 
		arrows indicate the direction of field changes.  }
        \protect\label{X/M(H)41AuNb}
\end{figure}
\begin{figure}
\includegraphics[width=0.96\linewidth]{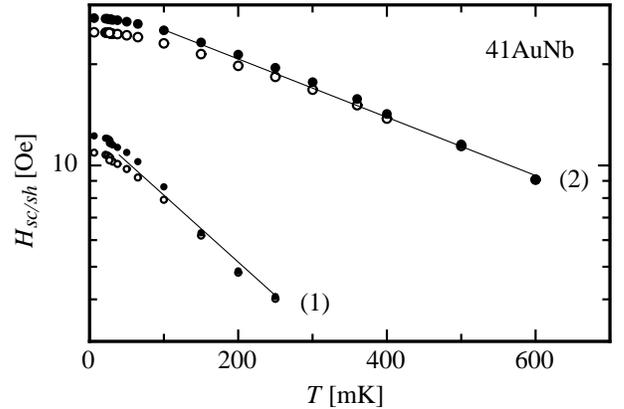}
		\caption{Supercooled ($\circ$) and superheated ($\bullet$) 
		field as a function of temperature and exponential fits [see 
		Eqn.\ \ref{hscsh}] for sample~41AuNb.  The different data 
		correspond to the low-(high-)field transitions for effective 
		gold thickness $d^{(1)}_{N}=4.0\,\mu\mathrm{m}$ 
		($d^{(2)}_{N}=1.8\,\mu\mathrm{m}$).  }
        \protect\label{Hb(T)41AuNb}
\end{figure}

For comparison, in Fig.\ \ref{X(logT)41/3} we also have plotted the 
temperature dependent diamagnetic susceptibility of a silver-niobium 
sample described in Ref.\ \onlinecite{bmueller2} [sample~3AgNb, 
$d_{N}=5.5\,\mu{\mathrm m}$, $T_{A}=310\,\mathrm mK$] and its 
numerical fit with $\ell/d_{N}=0.55$.  A close examination of both 
curves shows that, the diamagnetic susceptibility of the gold-niobium 
sample develops not so steep as for the AgNb sample, due to a higher 
level of impurities.  Moreover, as has been shown by the 
authors\cite{bmueller1} and in Fig.\ \ref{X(logT)41/3}, there are 
considerable deviations of the experiment from the theoretical curves 
at temperatures $T<T_{A}$, where the surface quality has a stronger 
influence, not accounted for in the theory\cite{belzig}.

We also have investigated the nonlinear-magnetic response in the 
low-temperature regime, where the induced diamagnetism shows 
approximately full Meissner screening.  The magnetic breakdown of the 
induced superconductivity, a first-order phase transition, was 
measured for the gold-coated NS proximity sample.  In Fig.\ 
\ref{X/M(H)41AuNb}(a), the isothermal nonlinear susceptibility as a 
function of magnetic field is shown for different temperatures between 
$7\,\mathrm{mK}$ and $400\,\mathrm{mK}$.  An isothermal 
magnetization curve for $T=7\,\mathrm{mK}$ is displayed in Fig.\ 
\ref{X/M(H)41AuNb}(b).

The double transitions displayed in Fig.\ \ref{X/M(H)41AuNb} are 
caused by the displacement of the niobium cores from the center of the 
cylinders during our drawing procedure of the niobium inside the soft 
gold matrix.  The two nearly separated transitions $(1)$ and $(2)$ 
correspond to two different effective thicknesses of the gold layer 
$d_{N}^{(1)}$ and $d_{N}^{(2)}$.  The supercooled and superheated 
fields $H_{sc,sh}^{(1,2)}$ for each breakdown transition are shown as 
a function of temperature in Fig.\ \ref{Hb(T)41AuNb}.  At temperatures 
well above $0.1T_{A}$, the experimental breakdown fields follow the 
exponential dependence
\begin{equation}
\label{hscsh}
H_{sc,sh}^{(1,2)}(T)\propto 
\frac{1}{d_{N}^{(1,2)}}\exp\left[-d_{N}^{(1,2)}/\xi_{N}(T)\right]\, ,
\end{equation}
indicated in Fig.\ \ref{Hb(T)41AuNb} as solid lines.  This is similar 
to the theoretical clean limit expression as given in Ref.\ 
\onlinecite{fauchere}.

Experimentally we found 
\begin{equation}
\xi_{N}(T)=p\cdot \xi_{T}\, , 
\end{equation}
with the clean limit thermal length in gold 
\begin{equation}
\xi_{T}=\hbar v_{F}/2\pi 
k_{B} T=1.7\,\mu\mathrm{m}/T(\mathrm{K})\, .
\end{equation}
For this gold-niobium sample~41AuNb, the value of the prefactor $p$ is 
$0.5$.  Several other gold-niobium specimens showing paramagnetic 
reentrance have $0.4<p<0.7$.  The temperature dependence of Eqn.\ 
\ref{hscsh} with $p<1$ is analogous to our findings in silver-niobium 
samples with $0.15<\ell_{N}/d_{N}<0.3$ and copper 
samples\cite{journlowtempphys,bmueller1}, indicating an elevated 
density of scattering centers with respect to the clean samples with 
$p=1$.
\begin{figure}
\includegraphics[width=0.96\linewidth]{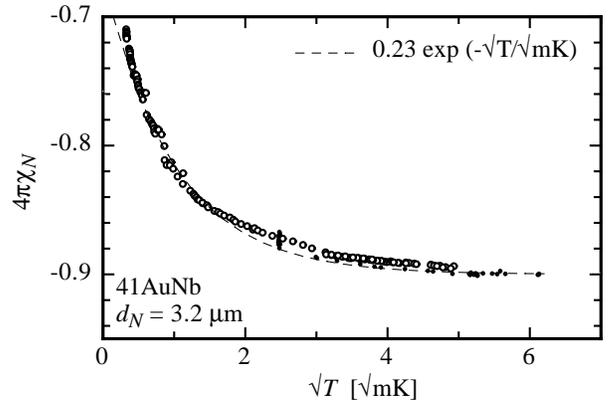}
		 \caption{Magnetic susceptibility $\chi_{dc}(T)$ at 
		 $H_{dc}=0.2\,\mathrm{Oe}$ ($\circ$) and $\chi_{ac}(T)$ 
		 ($\bullet$) below $T_{r}=40\,\mathrm mK$ as a function of 
		 square-root of temperature, for sample~41AuNb.  The dashed 
		 line is an exponential fit with $4\pi\chi_{\mathrm para}=0.23 
		 \exp{(-\sqrt{T}/\sqrt{1\,\mathrm mK})}$.  }
        \protect\label{X(sqrtT)41}
\end{figure}

In the linear magnetic susceptibility [Fig.\ \ref{X(logT)41/3}], for 
sample~41AuNb below $T_{r}\sim 40\,{\mathrm mK}$ the signature of 
reentrance is observed, with the development of an additional 
paramagnetic susceptibility $\chi_{\mathrm para}(T)$, such that 
$\chi_{N}(T)=\chi_{\mathrm dia}(T)+\chi_{\mathrm para}(T)$.

For this sample, the paramagnetic reentrance occurs at lower 
temperatures than for AgNb samples of similar sizes and shows a weaker 
increase as the temperature is reduced.  In spite of the smaller value 
of $\chi_{\mathrm para}$ of the gold-niobium sample, the reentrant 
effect below $\approx 500\,\mu\mathrm K$ shows a strong temperature 
dependence with no signs of saturation down to the minimal temperature 
$\approx 100\,\mu\mathrm K$.  The silver-niobium sample in Fig.\ 
\ref{X(logT)41/3} shows saturation of the magnetic susceptibility 
below $\approx 400\,\mu\mathrm K$, which could be intrinsic or lack of 
equilibrium.  Indeed, long-time effects of the order of several days 
showing for the silver-niobium samples discussed in Ref.\ 
\onlinecite{bmueller2}, have not been observed in sample~41AuNb.

Hysteresis effects which were very important in the silver-niobium 
samples described in Ref.\ \onlinecite{bmueller2}, were not observed 
in the gold-niobium sample, possibly due to the very small value of 
$\chi_{\mathrm para}$ above $7\,{\mathrm mK}$, where we could trace it 
on cooling and warming.

Fig.\ \ref{X(sqrtT)41} shows the reentrant paramagnetic susceptibility 
below $T_{r}$.  The data exponentially increases as
\begin{equation}
	4\pi\chi_{\mathrm para}(T)=A\exp{(-\sqrt{T/T_0})}\, ,
\end{equation}
with characteristic temperature $T_{0}=1\,\mathrm mK$ and prefactor 
$A=0.23$.  The value of the characteristic temperature can be 
connected to an energy through $k_{B}T_{0}=4.7\hbar v_{F} l_{N}/6\pi 
L^2$, with $l_{N}$ the measured mean-free path and $L=72\,\mu\mathrm 
m$ the wire perimeter.

Alternatively, we use the experimental coherence length of the Andreev 
pairs $\xi_{N}=p\xi_{T}$ in Au, obtained from our breakdown field 
measurements, in the expression for the dirty limit coherence length 
$\xi^d_{N}=\sqrt{1/3\xi_{N}\ell_{N}}$.  Thus, we find for the 
temperature dependence of the paramagnetic reentrant susceptibility
\begin{equation}
	\label{chipara}
4\pi\chi_{\mathrm para}=A \exp{\left[-\frac{L}{3\xi^d_{N}(T)}\right]}\, .
\end{equation}

The dirty-limit coherence length $\xi^d_{N}(T)$ in Eqn.\ \ref{chipara} 
again indicates the stronger level of impurities in the gold-niobium 
sample~41AuNb as compared to the silver-niobium samples discussed in 
Ref.\ \onlinecite{bmueller2}.  Due to the short mean-free path 
$\ell_{N}$ the induced diamagnetism is only slightly weaker as 
compared to sample~3AgNb, whereas the paramagnetic reentrant 
susceptibility is in the dirty regime.  $\chi_{\mathrm para}$ develops 
below a lower characteristic temperature $T_{r}$ and is by a factor of 
five smaller than for sample~3AgNb.  From these results it is evident 
that, impurities have a much stronger influence on the paramagnetic 
reentrance phenomenon than on the proximity effect.

In summary, we have found the paramagnetic reentrance phenomenon in 
gold-coated niobium proximity cylinders.  For these NS samples, the 
absolute value of the paramagnetic reentrant susceptibility is 
considerably smaller than in clean silver-niobium specimens, showing a 
stronger influence of scattering centers, the quality of the NS 
interface, as well as the free surface.  The temperature dependence of 
$\chi_{\mathrm para}$ is governed by the dirty limit coherence length, 
which becomes of the order of the wire perimeter.

The paramagnetic reentrant behavior of gold--niobium cylinders has to 
be viewed in light of the expected superconductivity in Au below 
$T_{c}\approx 200\,\mu{\mathrm K}$ \cite{hoyt} for a very pure gold 
sample, as well as the theoretical assumptions\cite{fauchere99,maki} 
for this element.

We wish to express our gratitude to M. Anen and M. Nider{\"o}st, for 
their contributions to the experimental work.  We especially thank the 
group at the ULT facility in the University of Bayreuth for their help 
and support.  We acknowledge discussions with W. Belzig, G. Blatter, 
C. Bruder, K. Maki, and Y. Imry.  Partial support from the 
``Schweizerischer Nationalfonds zur F{\"o}rderung der 
Wissenschaftlichen Forschung'' and the ``Bundesamt f{\"u}r Bildung und 
Wissenschaft'' (EU Program ``Training and Mobility of Researchers'') 
is acknowledged.

\end{multicols}


\begin{thebibliography}{00}
\bibitem{flukiger} {The initial sample preparation was done in the 
lab of R. Fl{\"u}kiger at the University of Geneva.
}
\end{thebibliography}
\end{document}